\documentclass[twocolumn,aps,showpacs,showkeys]{revtex4}
\usepackage{graphics}% Include figure files
\usepackage{mathrsfs}
\usepackage{stmaryrd}
\usepackage{verbatim}
\begin{document}
\author{W.\ De Baere}
\email{willy.debaere@UGent.be}
\affiliation{Unit for Subatomic and Radiation Physics,
Laboratory for Theoretical
Physics, State University of Ghent, Proeftuinstraat 86, B--9000 Ghent,
Belgium}

\newcommand{\cs}{\cal S}
\newcommand{\hl}{\lambda}
\newcommand{\ket}[1]{\left|#1 \right>}
\newcommand{\mv}[1]{\left< #1 \right>}
\newcommand{\one}{\mbox{{\sf 1}\hspace{-0.20em}{\rm l}}}
\newcommand{\lraw}{\longrightarrow}
\newcommand{\llraw}{\longlongrightarrow}
\newcommand{\Llraw}{\Longleftrightarrow}
\newcommand{\Lraw}{\Longrightarrow}\vspace*{1.0cm}
\newcommand{\law}{\leftarrow}\noindent
\newcommand{\llaw}{\longleftarrow}{\large
\title{Renninger's Thought Experiment: Implications for Quantum
Ontology  and for Quantum Mechanics' Interpretation}

\begin{abstract}
It is argued that the conclusions obtained by Renninger
({\em Z.\ Physik} {\bf 136}, 251 (1953)), by means of an
interferometer thought experiment, have important
implications for a number of still ongoing discussions
about quantum mechanics (QM). To these belong the
ontology underlying QM, Bohr's complementarity principle,
the significance of QM's wave function, the ``elements of
reality'' introduced by Einstein, Podolsky and Rosen
(EPR), and Bohm's version of QM (BQM). A slightly
extended setup is used to make a physical prediction at
variance with the mathematical prediction of QM. An
english translation of Renninger's paper, which was
originally published in german language, follows the
present paper. This should facilitate access to that
remarkable, apparently overlooked and forgotten, paper.
\end{abstract}

\vspace*{.3cm}
\noindent

\maketitle

\section{Some historical notes}
Some 80 years ago the main equations of QM have been
invented, with a subsequent overwhelming succes of its
predictive power, see e.g.\ M.\ Jammer \cite{jammer66}.
In contrast, however, its significance or interpretation
is still the subject of intense debate. These
interpretations range between two possible extremes. One
extreme contains the Copenhagen--like interpretations
\cite{stappciqm}, to the other extreme belong the
``realistic'' interpretations. Whereas the former are
concerned mainly with relationships between measurement
outcomes and carefully avoid ontological statements, the
latter consider observations made by human observers as
properties {\em possessed} by real existing objects. Each
of these interpretations has its range of supporters,
and each claims to give a acceptable explanation of what
QM is really about.

An extensive historical survey  is given in M.\ Jammer's
classic book ``The Philosophy of Quantum Mechanics -- The
Interpretations of Quantum Mechanics in Historical
Perspective'' \cite{jammer74}. However, in the literature
the question is rarely addressed whether there exist
empirical data or thought experiments which could rule
out some of these interpretations. More precisely, the
issue is whether there exist unavoidable ontological
truths which should, therefore, be part of {\em any}
acceptable interpretation. The answer -- maybe unexpected
and surprinsingly -- is that such a truth exists. Indeed,
in 1953 M.\ Renninger published a paper ``Zum
Wellen--Korspuskel--Dualismus'' (``On Wave--Particle
Duality'') in {\em Zeitschrift f\"ur Physik}
\cite{renninger} in which an interferometer thought
experiment played a central role. The basic result of
Renninger was that, independently of any theory, physical
reality at the quantum level exists of extended objects
which {\em at the same time}, i.e.\ in the same
experiment, have a wavelike {\em and} a particle--like
behaviour. Because this conclusion rests purely on
empirical facts, Renninger argues that it is compelling
and unavoidable. It should, therefore, be part of any
reasonable interpretation of QM. Yet even today, more
than 50 years after Renninger's paper, this is not the
case. How can this be, how is it possible that such a
fundamental truth has been overlooked and apparently
completely forgotten? One exception is M.\ Jammer's book
\cite{jammer74} where Renninger's paper is mentioned (p.\
494), together with some reactions by A.\ Einstein, M.\
Born, and P.\ Jordan. According to M.\ Jammer it ``caused
quite a stir'' among the experts in quantum physics.

It is interesting to recall that this picture of physical
reality was already introduced by L.\ de Broglie as early
as the introduction of the quantum formalism. It was
supported by e.g.\ A.\ Einstein and E.\ Schr\"odinger,
and later on used by D.\ Bohm in his attempt to set up an
alternative formulation of QM. For this reason we will
call this model the de Broglie--Einstein--Bohm (dBEB)
model. So, basically it was L.\ de Broglie who introduced
the idea that in reality, a quantum system should be
considered as an extended structure having at the same
time wave--like and particle--like properties. These
particle--like properties should then be characteristics
of a more localized region within the extended
phenomenon. In the dBEB model, in some way the localized
region -- or the ``particle'' -- is guided by the more
extended structure. In Bohm's version of QM, a so--called
``quantum potential'' is introduced to do this job. Now,
again invoking Renninger's ingenious analysis, the idea
of such a guidance is supported convincingly when the
extended structure moves through a system, such as a
Mach--Zehnder interferometer (MZI). Indeed, depending on
the wave properties of (various components of) that
structure as it moves through MZI, the subsequent
observation may occur in one of physically separated
detectors, and is the result of the interference of
different real waves (see Section 2).

From the lack of referring to Renninger's paper in later,
more recent, works on the interpretation of QM and, in
particular, on the ``reality of quantum waves'', it may
be concluded that this work has been unnoticed entirely
by the english and the french speaking physics community.
Indeed, it is never cited, neither by e.g.\ de Broglie
when -- after early criticism by W.\ Pauli -- he took up
again his idea, nor by Bohm while developing (together
with J.P.\ Vigier) his alternative approach to QM, which
was precisely based on de Broglie's model of a quantum
system. As a result, the fundamental importance of
Renninger's conclusions -- the existence of a causally
influencable ontological reality, in particular of the
reality of the so--called ``empty wave'' (EW) -- have
been overlooked and forgotten also by present day
physicists working in the field of the foundations of QM.
However, as the present author is convinced of the
fundamental importance of Renninger's penetrating
analysis of wave--particle duality underlying QM, an
english translation has been made of the original german
version. It is hoped that in this way many investigators
in the field of the foundations of QM may reconsider or
revisit their ideas with respect to quantum ontology, the
significance of the quantum formalism, and other related
quantum issues (see also Section 3).

More recently, the following investigations were carried
through to prove the reality of EWs in the sense of
observably influencing other physical systems. First,
from 1980 on, there were studies by Selleri
\cite{selleriewfop82,selleriewanf82,selleriewlnc83},
Croca \cite{croca87} and others \cite{crocafopl90}. In
some of these proposals it was maintained that there was
an observable difference between QM and a theory based
from the outset on the dBEB model of reality. However,
experimental results obtained by L.\ Mandel and coworkers
\cite{wang91} ``\ldots clearly contradict what is
expected on the basis of the de Broglie guided--wave
theory, but are in good agreement with the predictions of
standard quantum theory\ldots''. In a subsequent comment,
P.\ Holland and J.-P.\ Vigier \cite{holland91} replied
that Mandel's results did not invalidate the dBEB model
of reality itself. Finally, more recently there was an
attempt by L.\ Hardy \cite{hardyew} to prove the
observable reality of EWs, which was criticized in
various papers
\cite{pagonisew92,griffithsew93,zukowskiew93,dbew},
followed by replies by Hardy
\cite{hardyewonpag92,hardyewonzuk93}.

The purpose of the present paper  is threefold. First, in
Section 2 we recall the basic results obtained by M.\
Renninger already in 1953 \cite{renninger} about the
ontology underlying QM, and emphasize {\em the
unavoidable character} of Renninger's conclusions. Next,
in Section 3, we give a brief survey of the implications
for some present day quantum issues. Finally, by
considering an alternative setup in Section 4, we extend
and complete Renninger's argumentation by showing that
empty waves not only are causally influencable (as proven
in \cite{renninger}), but are themselves able to
influence observably other physical systems. The main
difference with Hardy's argumentation is that we get our
conclusion {\em without the introduction of any specific
supplementary hypothesis} in order to ascertain the path
along which the ``particle'' possibly moves.

\section{Renninger's argumentation: main results}
Renninger's setup is essentially a  Mach--Zehnder
interferometer which we present here as in Fig.\ 1, in
which the same notations as in Renninger's original work
are used.

A source in path $1$ prepares single  quantum systems
$\cal S$ (photons in \cite{renninger}) which move towards
a lossles 50--50 beam splitter situated in a location 2.
From location 2, two paths lead to mirrors $S_A,S_B$, who
reflect the beams in path 6 and 7 to a second beam
splitter in location 3. Finally, detectors $D_1,D_2$ are
located in outgoing paths 4, 5. As a first essential
point, Renninger remarks that in his argumentation only
{\em empirical} facts are considered, and that, hence,
his conclusions are {\em independent} of any existing
theory used to explain these empirical facts. The
empirical facts considered by Renninger are then the
following ones:
\begin{figure}
\centering
\scalebox{0.9}{\includegraphics{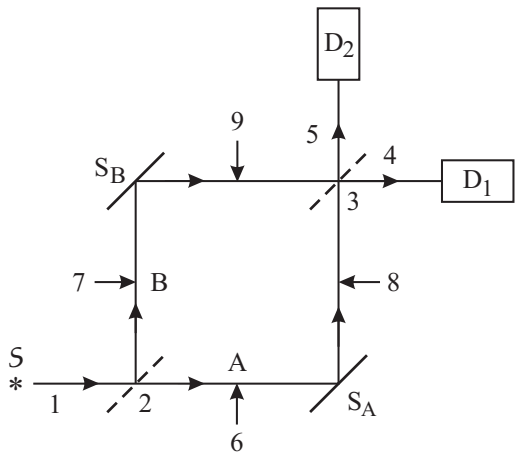}}
\caption{Renninger's interference set--up}
\end{figure}
\begin{itemize}
\item[a.]
If detectors are placed in paths 6-8 or 7-9, then
detection will occur in only one path at the same time,
never in both together.

This proves the {\em particle--like} aspect of the
phenomenon of the passage of the quantum system through
MZI.
\item[b.]
For equal path lenghts 6-8 and 7-9, with each single
quantum system moving through MZI, there will correspond
an observation with certainty in $D_1$, while nothing
will be observed in $D_2$.

This may be interpreted as the result of interference of
waves simultaneously moving along paths 6--8 and 7--9,
and should be considered as evidence for the {\em
wave-like} aspect of the phenomenon {\em in the same
setup}.

This has been verified by Grangier {\em et al.}
\cite{aspect861} for light.
\item[c.]
If a transparent ``half--wave'' plate is inserted at a
specific location in one of the paths 6-8 or 7-9 at
appropriate times (i.e.\ before the system $\cal S$ has
passed that location), then observation behind a beam
splitter at 3, will now occur with certainty in detector
$D_2$.

The fact that it does not matter in which path the
half--wave plate is inserted, is according to Renninger
empirical evidence for the simultaneous motion -- or
existence -- of physical realities in both paths. Because
of a.\ above, Renninger speaks of an ``empty wave''
moving along one path, and of another wave containing the
``particle'' along the other path. In order to
distinguish furtheron both waves, we will call this wave
the ``full wave'' (FW), i.e.\ the one responsible for
transfer of particle--like properties. If necessary, we
will add an index $\cal S$, e.g.\ $\text{EW}_{\cal S}$,
$\text{FW}_{\cal S}$.
\item[d.]
The result of the action in c., i.e.\ steering  detection
from $D_1$ to $D_2$, may be suspended by inserting on a
later instant another -- or even the same! -- half--wave
plate in the same or in the other path. And, if the paths
are long enough, then the previous action may be repeated
an arbitrary number of times, each time in such a way
that it will be {\em causally predictable}, i.e.\ with
certainty, in which detector $D_1$ or $D_2$ observation
will take place.

This is further evidence for the reality of both EW and FW.
\end{itemize}
For the subtleties of the argumentation, we refer  to
Renninger's paper \cite{renninger} or to its translation
in english.

\section{Implications for various issues in
quantum mechanics}
\subsection{The nature of quantum systems}
It  follows from Renninger's thought experiment that
ontologically a quantum system is an extended structure
consisting of realities which, under appropriate
circumstances (such as after the passage through a beam
splitter), may move along paths largely separated in
space. Hence, in considering physical processes it is
reasonable to make the hypothesis that, ultimately, it
are the properties of the {\em entire} structure that
brings about a definite localized measurement outcome.
One such property is, e.g., the phase {\em difference}
between $\text{EW}_{\cal S}$ and $\text{FW}_{\cal S}$.

What Renninger has shown also is that, in one and the
same single experiment, {\em both} $\text{EW}_{\cal S}$
and $\text{FW}_{\cal S}$ may be influenced {\em in a
causal way} by placing another system ${\cal S}'$, in
particular $\text{FW}_{\lambda/2}$, in either path 6--8
or 7--9. Here the causal character of the influence
refers to the certainty of predictions about observations
after a subsequent interaction with the second  beam
splitter in the location 3.

Also, it might be  the case that both ontological
components of a quantum system $\cal S$ -- i.e.\
$\text{EW}_{\cal S}$ and $\text{FW}_{\cal S}$ -- may be
influenced either separately or both at the same time,
again in a causal way in the case of a
$\lambda/2$--system. And furthermore, the possibility
should be envisaged that, conversely, both realities
themselves have the ability to influence observably the
realities of other physical systems. In particular this
would imply that $\text{EW}_{{\cal S}'}$ associated with
one system ${\cal S}'$ should be able to influence
$\text{EW}_{\cal S}$ associated with another system $\cal
S$, changing in this way the wave guiding $\cal S$, which
finally gives rise to a localized observation. This
possibility is discussed in Section 4.

So, Renninger's thought experiment has revealed that
ontologically -- and independent of any physical theory
-- quantum systems should be considered {\em at the same
time} as consisting of an {\em extended structure} with
wavelike properties, {\em and} a more localized region
within this structure which is able to exchange with
other systems, properties which are characterized by
means of particle--like variables such as energy $E$,
momentum $p$, spin, etc. Basically, Renninger's work
confirms -- on empirical grounds -- the validity of the
dBEB picture of physical reality.

\subsection{Bohr's complementarity}
According to Bohr's complementarity  principle, the
behaviour of a quantum system in a particular proces is
either wave--like or particle--like. In essence, it is
the whole setup, including the measurement apparatus,
which determines whether a quantum system behaves as a
particle or as a wave, but never in both ways together.
However, Renninger's interferometer thought experiment
clearly shows that the passage of {\em one single}
quantum system through the simplest version of a MZI
reveals both aspects at the same time: a detector placed
in either of the paths 6--8 or 7--9 reveals its
particle--like behaviour, while the causal effect of the
insertion of a $\lambda/2$ plate in either path -- and
the subsequent observation in either $D_1$ or $D_2$ --
should be interpreted as evidence of its wave--like
behaviour.

It is interesting to note here that Renninger's
empirically based conclusion predates by 40 years similar
conclusions obtained by e.g.\ Ghose and D.\ Home
\cite{ghosehome93,ghosehome96}.

\subsection{The significance of the wave function}
Like EPR, it was not Renninger's aim to criticize QM, but
only ``\ldots to point to some very precise conclusions,
which follow merely from purely experimental physical
aspects, without any previous knowledge of the
mathematical quantum formalism\ldots''.  Once these
conclusions have arrived at, Renninger is, however, very
clear about the significance of QM's mathematical
formalism: ``Of course one is free, to speak of the wave
as a pure ``probability''--wave. But one should be aware
of the fact, that this probability wave propagates in
space and time in a continuous way, and in a way that she
can be influenced in a finite region of space -- and only
there! -- and also at that time! --, with an unambiguous
observable physical effect!''

So, by Renninger's result the meaning  of the QM
wavefunction $\psi$ is very clear, both ontologically and
mathematically: ontologically $\psi$ represents a real,
causally influencable, wave, and mathematically it
satisfies the deterministic Schr\"odinger equation. The
important significance of Renninger's analysis is to have
revealed, in a compelling and unavoidable way, the
existence of a deeper lying layer of reality, the causal
and quantitative behaviour of which is mathematically
described by the standard quantum formalism. Therefore,
it is fair to say that Renninger's results should be part
of any acceptable interpretation of QM, and that it is
unreasonable to discard quantum ontology, and look at QM
only in a pure mathematical way. Hence, it is incorrect
to claim that the meaning of $\psi$ is nothing more than
a mathematical {\em function} which enables one to
predict future statistical results for given initial
conditions. Yet, this viewpoint has been defended from
the early days of QM (e.g.\ by advocates of Bohr,
Heisenberg and others), thereby neglecting Renninger's
findings, right up to now (see e.g.\ the provocative
paper by Peres and Fuchs, ``Quantum theory needs no
`Interpretation''' \cite{fuchsper}), probably because of
being innocent of Renninger's basic work.

Therefore, I think that Renninger's conclusions answer
many of today's issues with respect to the significance
of the wave function. And, had the significance of
Renninger's work been appreciated properly in past and
recent times, it would have influenced significantly many
other papers. In fact, the overwhelming amount of
relevant literature should have reduced considerably.
Therefore, it is pointless to make a selection from among
the huge list of available references -- anyone should
make his own selection, and judge in what sense the
papers' statements should be adjusted by taking into
account Renninger's 1953 analysis.

\subsection{The issue of Einstein locality}
In his paper Renninger strongly argues in favour of the
validity of the locality principle (``Einstein
locality'') underlying Einstein's successful relativity
theories. In his discussion on a possible alternative for
the ontological reality of the EW, he states that the
only alternative for explaining the wavelike behaviour of
a quantum system, e.g.\ in a MZI setup, would be the
introduction of a ``normal electromagnetic wave'' with
the unavoidable consequence that ``\ldots at the moment
of absorption the wave would contract with superluminal
speed, and moreover through closed walls. {\em Such
assumption would be completely unacceptable}'' [emphasis
by the present author]. Of course, at present one could
object that Renninger's analysis predates about one
decade Bell's investigations \cite{bell64} of the EPR
issue, and that according to the present majority view,
Bell's theorem ``proves'' a nonlocality property of QM.
However, in recent and past work (see e.g.\
\cite{dbfop051}) we have argued strongly in favour of the
validity of Einstein locality, and concluded that the
breakdown of counterfactual definiteness at the level of
individual quantum processes would be a far more
reasonable explanation for the violation of Bell's
inequality by QM, and for resolving all other so--called
contradictory results.

\subsection{EPR's elements of physical reality}
In his paper, Renninger also reports about  Einstein's
interest in his analysis. In particular, EPR's notion of
``elements of physical reality'' is mentioned as having
inspired Renninger's own definition of ``physical
reality'' (see the Abstract of \cite{renninger}): ``The
notion `physical reality' should be understood such that,
when this physical reality is considered in a particular
space at a particular time, it should be experimentally
possible to influence this reality in such a way that
future results of experiments show unambiguously that
this reality has been causally influenced by the
experimental act in this space and at that time.''

Here one may remark that, as in the EPR paper, Einstein,
there is a clear relationship
between predictions for later observations in some
detector, and formerly existing, causally influencable,
realities. Because of this relationship one may identify
Renninger's realities with EPR's definition of ``elements
of physical reality''. In this sense, the realities
corresponding with EWs may be considered as EPR
``elements of physical reality'', of which one might
reasonably expect, as claimed by EPR, that they have a
representation in the physical theory. However, according
to Renninger, this is not the case with the present QM
formalism: ``\ldots the proven reality of the wave
associated with the single particle, which quantum
mechanics, \ldots, is unable to account for \ldots''.

\section{On Bohm's Version of Quantum Mechanics}
Because Renninger's analysis confirms on empirical
grounds the dBEB picture of reality at the quantum level,
one might be tempted to conclude that Bohm's
reformulation of QM in terms of the notion of ``quantum
potential'' supersedes the standard quantum formalism.
However, one of Bohm's intentions was to present a causal
quantitative formalism for the description of {\em
single} quantum processes, and to get back immediately --
almost {\em by construction} -- the statistical
predictions of QM.

As is well known \cite{hollandbook}, the price to be
payed  was that the quantum potential should be
interpreted as resulting from an instantaneous
action--at--a--distance between quantum systems, and with
an intensity which is independent of the mutual distance
between these systems. Here the term ``quantum system''
should be understood as that localized part of the
extended structure which has the characteristic to
exchange particle--like properties such as $E,p,$ etc.\
with other systems.

However, it is precisely because of the conflict between
the explicit nonlocality property of Bohm's quantum
potential with the empirical validity of locality in
actual physical processes, that BQM cannot be considered
as a valid quantitative scheme for {\em individual}
quantum processes. Recalling that QM in general makes
deterministic predictions for the statistics of
measurement outcomes, the minimum requirement for any
attempt to reproduce these predictions in terms of a {\em
local} theory for {\em individual} processes should be
the introduction of extra or supplementary variables in
the theory. This means that BQM cannot be considered as
such a valid alternative HVT for QM. At best, it is only
a {\em reformulation} of QM in terms of another
mathematical quantity -- the quantum potential -- which
should be considered only from a mathematical point of
view, i.e.\ it may not be considered as a faithful
representation of the underlying physical reality lead
bare by Renninger.

Probably, this was one reason for Einstein to consider
Bohm's solution as ``too cheap''(see e.g.\
\cite{bohrsolvay}). Therefore, an acceptable alternative
would be a theory based on the dBEB picture in which the
validity of locality is retained. However, not only does
such a formalism not yet exist, but neither is there any
onset to set up such a formalism (see, however,
\cite{hesspnas01b}).

\section{Another possible physical consequence of the
$\text{d}$BEB picture of reality}
\subsection{Physical argumentation}
In his work, on page 9, Renninger addresses the following
question: ``What happens to the wave devoid of energy of
a photon after its absorption? When it is absorbed for
example in (6), and when in addition the detectors in (4)
and (5) are removed, what happens then with the wave in
$B$? Does she move further towards infinity, or does she
disappear at the moment of absorption? Of course this
question cannot be answered principally. The former
assumption appears to be the more natural one, because it
avoids the conclusion to the existence of influences
which propagate with infinite speed also through closed
walls, a conclusion which within the physical world is
inconceivable. In any case were such influences not
associated with transport of energy.''

Although Renninger is of the opinion that ``\ldots this
question cannot be answered principally\ldots'', we yet
propose a slightly modified setup in which this issue
might be clarified. In fact, we follow Renninger's own
proposal that the dBEB model of reality can be used as
``a valuable aid for the visual comprehension of
elementary processes and for making exact prognoses about
the outcome of experiments.'' Because the modified setup
is almost as simple and elementary as Renninger's setup,
we have some confidence about the correctness of our
predictions applying to the new setup. These predictions
now concern observable effects of the EW reality on other
physical systems and, as in Renninger's analysis, only
physical arguments are invoked.

In this section we elaborate  on the ontological picture
resulting from Renninger's argumentation. If the
behaviour of a quantum system is determined by its
ontological constitution -- consisting of {\em both} wave
and particle characteristics -- one should be able to
influence this behaviour either by changing the wave
characteristics, or by changing the particle
characteristics. The first possibility has been evidenced
by Renninger's point c.\ in Section 2, i.e.\ the
insertion of a {\em material} system -- a half--wave
plate -- in either path. Hence, in this case it is the
component $\text{FW}_{\lambda/2}$ which provokes the
observable influence. The question then naturally arises,
whether the wavelike properties of the components
$\text{EW}_{\cal S}$ or $\text{FW}_{\cal S}$ of $\cal S$
may also be influenced by the empty wave component of
another system, e.g.\ the component $\text{EW}_{{\cal
S}'}$ of quantum system ${\cal S}'$.

This implies that one should be able to observe the
effect of superposing the realities $\text{EW}_{{\cal
S}'}$ with $\text{EW}_{\cal S}$ or $\text{FW}_{\cal S}$.
To this end we supplement Renninger's setup in  a minimal
way by adding 1) another source of systems ${\cal S}'$,
2) a third beam splitter $\text{BS}_{2'}$ in location 2',
and 3) by letting mirror $S_B$ removable (see Fig.\ 2).

Systems ${\cal S}'$ move then along direction 1' towards
$BS_{2'}$. The outgoing arms are $A'$, which is in line
with path 9, and $B'$ which contains a detector $D_{2'}$.
For given masses $m_{\cal S}$ and $m_{{\cal S}'}$ we
choose the velocities $v_{\cal S}$ and $v_{{\cal S}'}$
such that the frequencies of the waves associated with
$\cal S$ and with ${\cal S}'$ are the same.
\begin{figure}[htp]
\center \scalebox{0.9}{\includegraphics{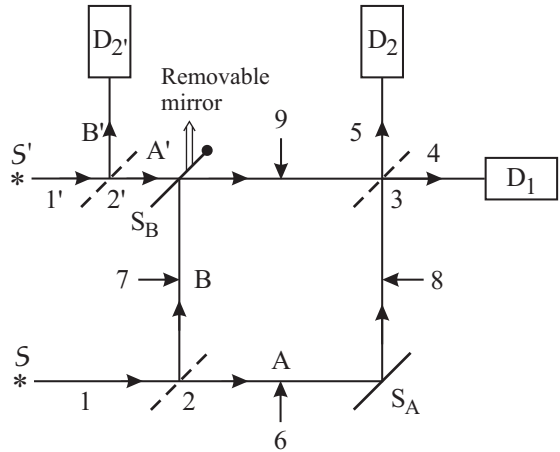}}
\caption{Modified
interference set--up}
\end{figure}
We are now interested in those cases  where ${\cal S}'$
is observed in $D_{2'}$. Then with certainty
$\text{EW}_{{\cal S}'}$ is moving towards $S_B$. We
assume further that the dimensions of the arrangement
allow the following: after $\text{FW}_{\cal S}$ or
$\text{EW}_{\cal S}$ has passed $S_B$, $S_B$ is removed
so that at a later instant $\text{EW}_{{\cal S}'}$ can
pass  freely, move further along path 9 and, finally,
catch up either $\text{FW}_{\cal S}$ or $\text{EW}_{\cal
S}$, which moved equally along path 7--9.

From Renninger's analysis we know  that steering the
future observation of $\cal S$ -- to $D_1$ or to $D_2$ --
cannot take place by installing a half--wave plate at
positions where $\text{EW}_{\cal S}$ or $\text{FW}_{\cal
S}$ has already passed. For the same reason one may
assume that, after $\text{EW}_{\cal S}$ or
$\text{FW}_{\cal S}$ has passed $S_B$, the removal of
$S_B$ does not have any influence -- or steering -- on
the future course of the processes in the MZI, in
particular it has no influence on the detector where
observation of $\cal S$ will take place. Assume then that
at such an appropriate instant the mirror $S_B$ is
removed, and that the various path lengths and velocities
of both $\cal S$ and ${\cal S}'$ are chosen in such a way
that $\text{EW}_{{\cal S}'}$ and the component of $\cal
S$ moving along path 9 (i.e.\ either $\text{EW}_{\cal S}$
or $\text{FW}_{\cal S}$), will superpose {\em just
before} entering $\text{BS}_3$ at 3 (see Fig.\ 3,
where only the last stage before entering $BS_3$ is
shown). As a result of this superposition the wave
properties of $\cal S$, in particular the phase, will
have been changed.
\begin{figure}[htp]
\center
\scalebox{0.7}
{\includegraphics{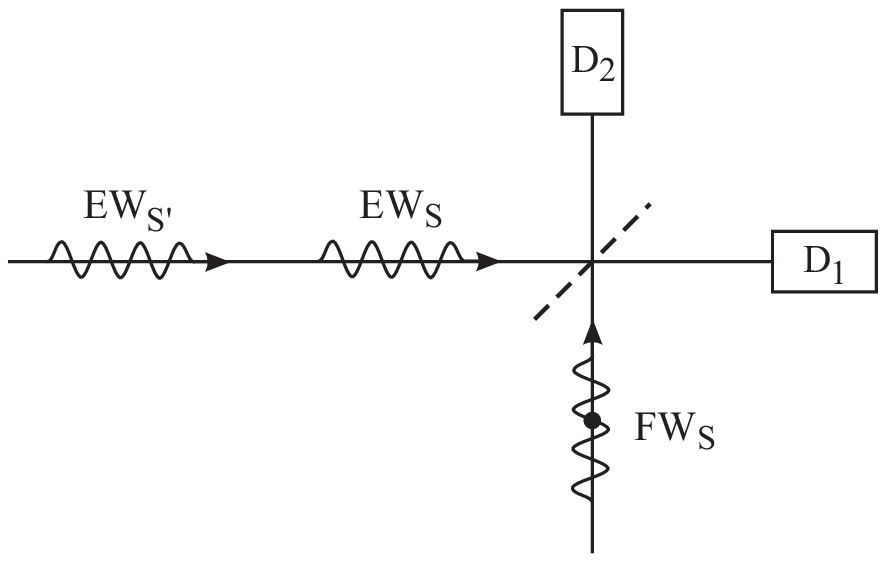}}\\[3mm]
\scalebox{0.7}
{\includegraphics{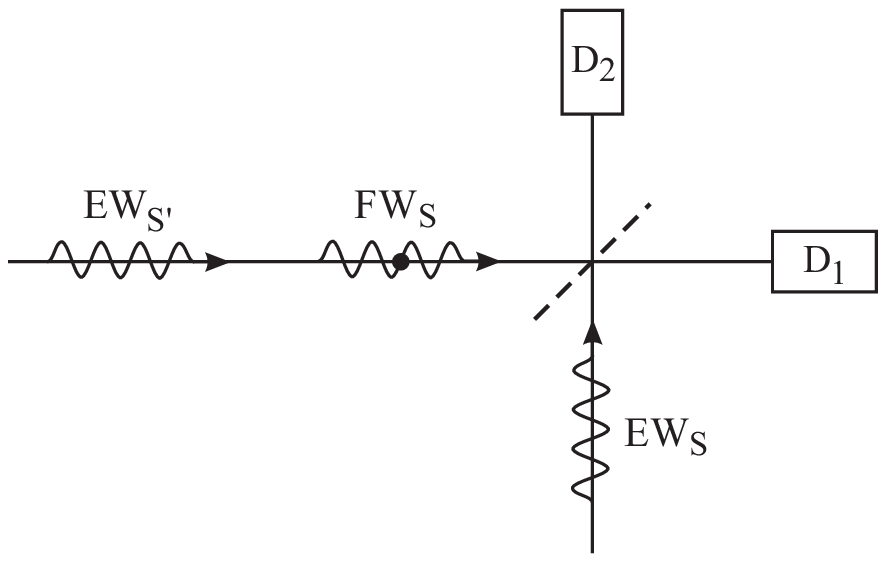}} \caption{Interaction
between $\cal S$ and $\text{EW}_{{\cal S}'}$}
\end{figure}
Then it might legitimately be expected that, as a result
of this interaction, the coherence  between
$\text{EW}_{\cal S}$ and $\text{FW}_{\cal S}$ has been
disturbed, and that the final superposition of these real
waves within $BS_3$ will no longer give rise to
observations {\em in detector $D_1$ only}. Observation of
$\cal S$ in $D_2$ should then be considered as evidence
for the ability of empty waves to influence other
physical systems, i.e.\ either another empty wave or a
material system, in a directly observable way. As a final
remark, we note that in the above reasoning we have only
used Renninger's conclusion of the reality of empty
waves, and supplemented it only by the reasonable
assumption that physical realities may be changed by
interactions among them.

\subsection{Quantummechanical predictions}
Let us now look at how QM describes the processes in this
minimally extended MZI setup. The systems leaving the
sources in paths 1 and 1' may be assumed independent, and
their state is the direct product $\ket{1}\ket{1'}$.
Passage of ${\cal S}'$ through $BS_{2'}$ and of $\cal S$
through $BS_2$ results in the evolution:
\begin{eqnarray}
\ket{1'}\ket{1} &\stackrel{BS_{2'},BS_2}{\lraw}
\frac{1}{\sqrt{2}}(\ket{A'}e^{ik'z_{A'}}
+i\ket{B'}e^{ik'z_{B'}})
\nonumber\\
&\times\frac{1}{\sqrt{2}}
(\ket{A}e^{ikz_{A}}+i\ket{B}e^{ikz_{B}}), \label{eendoorbs}
\end{eqnarray}
where $\ket{A}$ represents the state along a horizontal
path, $\ket{B}$ the state along a vertical path, etc.
Now, because we will be interested in observations in
$D_1$ and $D_2$ for equal path lenghts 6--8 and 7--9, we
may replace already at this stage the lengths $z_A$ and
$z_B$ by $z$. In a similar way we replace $z_{A'}$ and
$z_{B'}$ by $z'$. This gives rise to common exponential
factors $e^{ikz}$ and $e^{ik'z'}$. Reflection at the
mirrors amounts to a phase shift of the states by
$\frac{\pi}{2}$, and the QM state transforms further to
\begin{equation}
\stackrel{S_A,S_B}{\lraw}
\frac{1}{\sqrt{2}}(\ket{A'}+i\ket{B'})e^{ik'z'}
\frac{1}{\sqrt{2}}(i\ket{B}-\ket{A})e^{ikz},
\label{eendoorbs1}
\end{equation}
At this moment, observation  in detector  $D_{2'}$ is
recorded, and only those cases where ${\cal S}'$ is
observed in $D_{2'}$ are  retained. This subensemble is
described quantummechanically by the state
\begin{equation}
\ket{\Psi_{12}}= \frac{1}{2} \ket{A'}e^{ik'z'}
(i\ket{B}-\ket{A})e^{ikz}.
\end{equation}
It follows that the systems $\cal S$ in this subensemble
are still described by the QM state
\begin{equation}
\ket{\Psi_{\cal S}}=\frac{1}{2}(i\ket{B}-\ket{A})e^{ikz}
\end{equation}
Finally, after the passage through $BS_3$, this
state becomes again:
\begin{equation}
\ket{\Psi_{\cal S}}= i\frac{1}{2}(\ket{B}+i\ket{A}) e^{ikz}
\stackrel{BS_3}{\lraw}
-\frac{1}{\sqrt{2}}\ket{A}e^{ikz},
\end{equation}
so that QM predicts that all systems $\cal S$ will still
be observed in detector $D_1$. Clearly this reflects the
fact that the QM formalism predicts that EWs cannot
influence other quantum systems in an observable way.

So, observations in $D_2$ would  clearly be caused by the
influence of $\text{EW}_{{\cal S}'}$ on system $\cal S$
itself.  This, then, would be conclusive evidence for the
possibility of EWs to influence in an {\em observable}
way other physical systems.

\section{Conclusions}
In this work we have reviewed  Renninger's penetrating
analysis leading to his empirical proof of the reality of
quantum waves, in particular  of the empty wave. We have
argued -- or, rather, called attention to Renninger's
opinion -- that these results have fundamental
ontological significance. If de Broglie should be
credited for the idea, then Renninger should certainly be
credited for the empirical proof of its validity. We also
discussed briefly the impact on some still ongoing issues
in the foundations of QM. In particular, the dBEB
ontological picture of reality should be part of any
acceptable interpretation of QM, implying
that many of these interpretations should be revised.

Next, we have proposed a slightly modified setup in
which, possibly, the real EW should influence -- instead
of being influenced by -- another quantum system in an
observable way. This influence should manifest itself by
an observation in detector $D_2$, whereas QM still
predicts no observation in that detector.

If it should turn out that QM gives the wrong prediction,
then a new formal scheme should be required for giving a
more faithful description of the EW reality -- a
description which QM is unable to give. Tentatively, this
could be realized by means of a {\em local} HVT having
the general characteristics described in \cite{dbfop051}.

\end{document}